\documentclass[%
 reprint,
 amsmath,amssymb,
 aps,
]{revtex4-1}

\usepackage{graphicx}% Include figure files
\usepackage{dcolumn}% Align table columns on decimal point
\usepackage{bm}% bold math
\begin{document}

%\preprint{APS/123-QED}
%'Dimension cross over' 

%near-diffusive -> normal
\title{Dimensionality-dependent crossover in motility of polyvalent burnt-bridges ratchets}% Force line breaks with \\

%Dimensionality-dependent crossover in motility of burnt-bridges ratchets

%Dimensional crossover from near-diffusive to ballistic motility of burnt-bridges ratchets

% % Transition from conventional to ballistic diffusion in burnt-bridges ratchets by dimensional crossover

% %Dimensional crossover from near-conventional diffusion to ballistic motility of burnt-bridges ratchets % %

%Brunt-bridges ratchets exhibit a dimensional-dependent crossover in  from near-diffusive to ballistic motility

\author{Chapin S. Korosec, Martin J. Zuckermann, and Nancy R. Forde}
\email{ckorosec@sfu.ca, mjzheb@gmail.com, nforde@sfu.ca}
 \affiliation{Department of Physics, Simon Fraser University, 8888 University Drive, Burnaby, British Columbia, V5A 1S6, Canada}%Lines break automatically or can be forced with \\

\date{\today}% It is always \today, today,

\begin{abstract}
The burnt-bridges ratchet (BBR) mechanism is a model for biased molecular motion whereby the construct destroys track binding sites as it progresses, and therefore acts as a diffusing forager, seeking new substrate sites. Using Monte Carlo simulations that implement the Gillespie algorithm, we investigate the kinetic characteristics of simple polyvalent BBRs as they move on tracks of increasing width. We find that as the track width is increased the BBRs remain nearly ballistic for considerable track widths proportional to the span (leg length) of the polyvalent walker, before transitioning to near-conventional diffusion on two-dimensional tracks. We find there exists a trade-off in BBR track association time and superdiffusivity in the BBR design parameter space of span, polyvalency and track width. Furthermore, we develop an analytical model to describe the ensemble-average motion on the track and find it is in good agreement with our Gillespie simulation results. This work offers insights into design criteria for \textit{de novo} BBRs and their associated tracks, where experimentalists seek to optimize directionality and track association time. 
\end{abstract}

%\keywords[showkeys]{Suggested keywords}%Use showkeys class option if keyword
\maketitle

%\tableofcontents

\section{\label{sec:level1}Introduction}

 %Unexpected track width dependent transitions in motility of burnt-bridges ratchets 

Diffusion, driven by random thermal motion, results in slow transport over long distances. Nature has overcome this problem through the evolution of impressive protein-based machines that achieve processive and directional motion despite their noisy thermal environment. Within the cell's cytoplasm, the molecular motors kinesin \cite{Svoboda1993}, dynein \cite{Nobutaka1998}, and myosin \cite{Mehta1999} achieve directional motion on their intracellular tracks by converting chemical energy in the form of ATP into mechanical stepwise translocation \cite{Schliwa2003, Bruno2009}. There are, however, other means by which cellular systems can achieve directed motion besides conventional cytoplasmic motors. In this work, we examine a class of machines that achieve directional motion by a `burnt-bridges ratchet'(BBR) mechanism. 

A BBR has a probability $p$ of destroying a substrate track site as it passes \cite{Antal2005}. Upon a successful cleavage event, the asymmetry produced in the track prevents backwards stepping. Motion forwards is driven purely by thermal motion without the need for an energetically driven conformational change in the walker. In order to achieve processive motion the timescale of track association must be long enough such that the BBR can cleave the substrate, explore neighbouring sites, and rely on thermal fluctuations to move. In one dimension, with $p=1$, the motion of a BBR is expected to be ballistic, while in two dimensions the motion is expected to resemble a self-avoiding walk. We also note that BBR nanomachines can be considered as diffusing foragers, where a parameter of interest is the number of cleavage events before the walker depletes its local environment and `starves' (detaches) \cite{Benichou2014}. 

 Matrix-metalloproteases (MMPs) are enzymes that move one-dimensionally via a BBR mechanism along collagen fibrils in the extracellular matrix \cite{Saffarian2004,Sarkar2012}. Individual MMPs have been observed to move superdiffusively along their collagen tracks at speeds up to $5.8\mu$m/s \cite{Collier2011}. In contrast to the one-dimensional motion of MMPs, the protein-based ParA/ParB system found in bacteria is an example of a two-dimensional BBR system \cite{Vecchiarelli2014a,Hu2017}. This system is responsible for partitioning extrachromosomal low-copy plasmid DNA during cell division \cite{Szardenings2010}. These BBRs have been observed to move directionally at speeds of $\sim0.1$ $\mathrm{\mu}$m/s on their two-dimensional tracks \cite{Vecchiarelli2014a}. Nature has therefore implemented the BBR mechanism in both one-dimensional and two-dimensional systems, where these BBRs have achieved speeds comparable to kinesin in saturating ATP conditions \cite{Block1990}. 

Inspiration from biological systems such as these has led to the development of synthetic nanomachines that achieve directional motion through various stepping mechanisms \cite{Yin2004,Pei2006,Bath2007,Bromley2009,Omabegho2009,Lund2010,Zhisong2012,Cha2014,Niman2014,Kovacic2015}. The motivation for the design and implementation of synthetic nanomachines is two-fold: to create a molecular system that mimics the behaviour of biological counterparts, thereby enabling us to learn about fundamental physical principles that give rise to observed biological molecular motor phenomena; and to create new technologies that perform tasks currently out of our reach \cite{Bath2007}. 

Many of the autonomous synthetic biologically-based nanomotors thus far realized are DNA-based BBRs \cite{Pei2006, Bath2007,Lund2010,Cha2014}. In the limit of low polyvalency, Cha et al. \cite{Cha2014} developed a DNA walker that moves in a self-avoiding fashion along carbon nanotubes by catalyzing cleavage of its RNA footholds. At the other extreme, DNA-coated microspheres, so-called `DNA monowheels', hybridize to a substrate surface coated with complementary RNA and have a high polyvalency with thousands of cleavable substrate contacts \cite{Yehl2015}. The DNA monowheel has demonstrated impressive velocities for an artificial system of up to 2 $\mu$m/min, as well as near-ballistic motion on its two-dimensional substrate track \cite{Yehl2015}. In contrast to the monowheel's high polyvalency, most DNA walkers are bipedal \cite{Omabegho2009, Liu2016, Li2018}. 

%\cite{Yehl2015}
%\cite{Omabegho2009, Liu2016, Li2018}

 In this work, we refer to the total number of legs as the polyvalency. Polyvalency of BBRs is thought to have a profound impact on  directionality and track attachment times \cite{Yehl2015}. Analytical approaches to understanding the effects of polyvalency on BBR dynamics are difficult as the memory requirement for visited sites leads to non-Markovian behaviour \cite{Antal2007a,Shtylla2015}. Because of this, researchers have largely turned to simulations to model the behaviour of synthetic BBR nanomotors \cite{Samii2010, Samii2011, Olah2013,Hu2015}.
 
Those who seek to experimentally develop synthetic nanomachines are met with the challenge of designing not only the the machine itself, but also the substrate track with which it is to interact. Missing from the literature is an exploration of how the width of the substrate track is expected to impact BBR kinetics. In this work we implement the Monte Carlo Gillespie algorithm \cite{Gillespie1977} to investigate the dependence of the mean squared displacement, track attachment time, kurtosis, and extent of substrate cleavage on the dimension of the substrate track. We generalize our results by altering the polyvalency and span of the BBRs to explore how these attributes influence ensemble-average kinetics of BBRs moving on tracks of increasing width. In this work we focus on ideal BBRs where substrate binding is followed by a probability $p = 1$ of catalyzing the bound site. Our BBRs cannot unbind from substrate without a cleavage event, and cannot rebind to a cleaved product site. Samii et al. \cite{Samii2011} report that nanomachines that can unbind from substrate and rebind to product display an increase in track attachment time as a function of increasing polyvalency. Similarly, Yehl et al. \cite{Yehl2015} report that the prolonged track attachment of their BBR DNA monowheel is because of the dramatically increased polyvalency. In contrast to these results, in our system we find that increasing the polyvalency of BBRs results in a dramatic decrease in track association time. Our results further indicate that reducing the dimensionality of the track to one dimension is not necessary to promote linear ballistic motion. There exists a tolerance window in track width that allows for maximally superdiffusive walkers.  

\section{\label{sec:level1}Model and methods}
\subsection{\label{sec:level2}Kinetic model}

\begin{figure}[h]
	\includegraphics[width=0.45\textwidth]{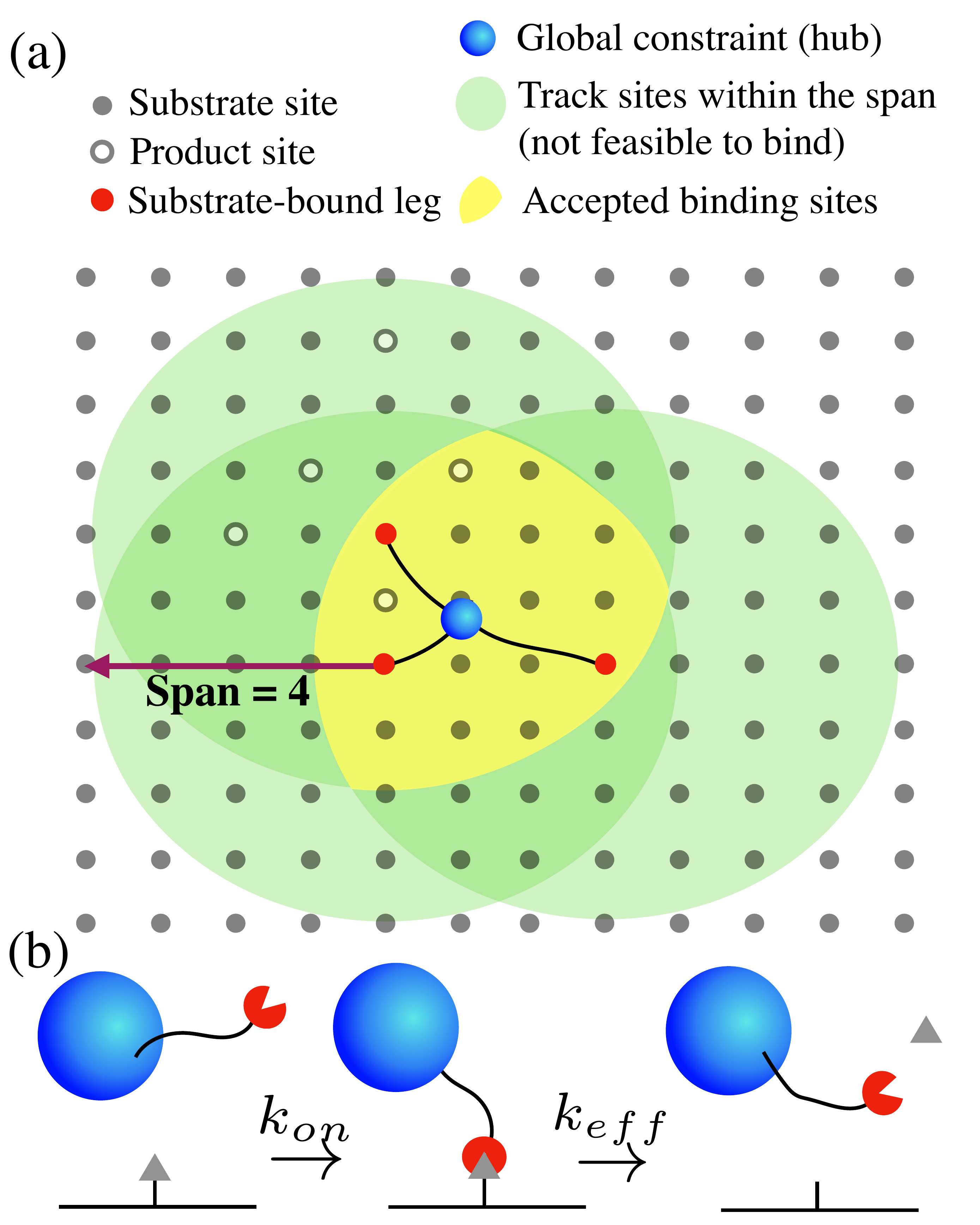}
	\caption{a) Each bound leg is assigned a span which defines the radius of a circle, shown in green, around the bound location, shown in red. The mutual overlap of bound-leg spans, shown in yellow, marks the feasible binding locations for unbound legs (not shown) during the following kinetic move. b) Each BBR leg (red) can interact with any availble substrate track site (grey triangle) via a rate of attachment, $k_{on}$. $k_{eff}$ includes both substrate cleavage and release.} 
	\label{trackSites}
\end{figure}

We model polyvalent BBRs as $n$ legs coupled to a point-like hub referred to as the \textit{global constraint} (Fig. \ref{trackSites}). The $n$ legs are non-interacting, but only a single leg can occupy any given track site. The legs chemically interact with the track via substrate binding and cleavage, followed by release from the cleaved product. To incorporate leg length into the kinetic model, each of the $n$ legs is assigned a $span$, which is defined as the maximum distance between any two bound legs. As shown in \textbf{Fig. \ref{trackSites}a} a circle of radius $R = span$ is drawn around each bound leg. The substrate track sites that fall within the mutual overlap of all legs' spans are considered binding options for the unbound legs. We note that our model is fundamentally different from that of Olah et al. \cite{Olah2013} where they allowed binding to all sites within a distance $\ell$ of each bound leg. We allow for binding within a small region around the global constraint, where the region (shown in yellow in \textbf{Fig. \ref{trackSites}a}) is determined by the currently bound legs. In this way we account for the collectively imposed constraint of all bound legs limiting the options for fresh track coupling.

 To study the motion of BBRs we developed a kinetic model similar to that used by Samii et al. \cite{Samii2010,Samii2011} and Olah et al. \cite{Olah2013} whereby we implement the Monte Carlo Gillespie algorithm \cite{Gillespie1977} to study polyvalent walker dynamics. Our kinetic model, as shown in \textbf{Fig. \ref{trackSites}b}, is a simple model that allows for substrate binding and substrate cleavage followed by unbinding. We employ a substrate binding rate, $k_{on} = 20 \; s^{-1}$, and cleavage rate, $k_{eff} = 0.054 \; s^{-1}$. $k_{eff}$ incorporates both the cleavage and detachment processes. These rates are similar to those used by Samii et al. \cite{Samii2010,Samii2011}. We made the decision to set the dissociation rates from uncleaved substrate sites to zero, and the product binding rate to zero. This allows us to to focus on a strict burnt-bridges ratchet system distinct from that of previous work on similar systems \cite{Samii2010,Samii2011,Olah2013}. Each substrate-bound leg is guaranteed to cleave and release to the unbound state where it may bind again to fresh substrate. Therefore, our legs have a probability of $p=1$ to cleave each bound substrate site. 
   
  A kinetic move is chosen by a Monte Carlo Gillespie algorithm that samples from all available transitions of all legs. For example, if there are 2 unbound legs, 3 bound legs, and 12 available substrate sites, there are 24 possible binding transitions and 3 possible transitions to cleave and release. A particular transition with rate $k_{i}$ is chosen with a probability $P_{i} = \frac{k_{i}}{\sum k_{i}}$. After a choice of transition is made, time is updated according to $t = \frac{1}{\sum k_{i}}\mathrm{ln}(\frac{1}{X})$, where $X$ is a random variable uniformly distributed on (0,1]. The central hub position is updated by determining the average position of the bound legs. We then track the motion of this point-like hub for kinetic analysis. 

\subsection{\label{sec:level2}Track design \& BBR parameters}

To explore the effect track dimensionality has on BBRs, we employed a large range of tracks that increase in width by factors of 2. In total we cover tracks of widths $2^{n}$ for $n = 0,1,2,3,...,12$, where a track width of $2^{0}$ is a one-dimensional track. For all 520,000 independent runs reported in this work, track widths of $2^{12} = 4096$ can be considered infinitely two-dimensional as no ratchet reached the boundaries within the maximum simulation time of 25,000 seconds for each independent run. As we varied the track width, the length of the track was consistently kept to 5000 lattice sites (effectively infinite in length). The track widths chosen were convenient as they allowed us to probe the effects of confinement through gradually increasing the track size away from one dimension into an infinite two-dimensional plane. All tracks were initialized as all-substrate tracks; we do not impose any initial asymmetry. All of the BBRs were initialized in the geometric centre of the track with one leg bound.

Motivated by previous work \cite{Pei2006,Samii2010,Lund2010,Samii2011,Olah2013,Vecchiarelli2014a,Kovacic2015,Hu2015,Yehl2015,Hu2017}, we focused on four BBR designs. In the notation of (polyvalency, span) we examined the behaviour of BBRs with parameters (12,8), (3,8), (12,3), and (3,3). The limit of (12,3) BBRs was chosen because all available binding locations within a span of 3 can be saturated. In this limit the BBRs are expected to produce a complete wake of cleaved sites such that the constructs cannot cross previously visited territory. This contrasts with (3,8) BBRs, which we expected to produce sparsely cleaved trajectories.

For each BBR design, on each track width, we ran 10,000 independent trajectories. For example, (3,3) has 10,000 independent runs on a track width of 1, and another 10,000 on a track width of 2, etc. Trajectories end when no BBR legs remain coupled to the track. We do not allow rebinding of detached BBRs. If the BBRs remain attached to the track for 25,000 seconds the simulation is also ended. 

%The limit of (12,3) BBRs was chosen because all available binding locations within a span of 3 can be saturated. In this limit the BBRs are expected to produce a complete wake of cleaved sites such that the constructs cannot cross previously visited territory, in contrast to (3,8) BBRs which are expected to produce sparsely cleaved trajectories.

\subsection{\label{sec:level2}Analytical methods}

\subsubsection{Mean squared displacement}

The mean squared displacement (MSD) is a useful measure used to assess the anomalous nature of a diffusive walk \cite{Havlin2002}. The MSD is defined as the variance in displacement, $\vec{X}$, and scales with a power-law dependence, 

\begin{equation}
\mathrm{MSD}(\vec{X}) \equiv \mathrm{Var}[\vec{X}] = \left\langle  (\vec{X} - \vec{\mu})^{2}\right\rangle \propto t^{\alpha}, 
\end{equation}
where $\vec{\mu}$ is the mean position and $\alpha$ describes the power-law scaling behaviour of the system. For a given ensemble of trajectories one can assess the slope of the log-log MSD-time plots to compute $\alpha$. For all BBR permutations and track widths we computed the MSD via ensemble averaging according to eq. 2, which is equivalent to eq. 1, 

\begin{equation}
\mathrm{MSD}(\vec{r}) = (\left\langle x^{2}\right\rangle - \left\langle x\right\rangle^{2}) + (\left\langle y^{2}\right\rangle - \left\langle y\right\rangle^{2}) \propto t^{\alpha_{r}},
\end{equation}
where $\vec{r}$ is the position of the BBR's global constraint. The $x$ and $y$ components of $\vec{r}$ can be examined independently. We therefore also define MSD($x$) and MSD($y$) as 

\begin{gather}
\mathrm{MSD}(x) = \left\langle x^{2}\right\rangle - \left\langle x\right\rangle^{2} \propto t^{\alpha_{x}},\\
\mathrm{MSD}(y) = \left\langle y^{2}\right\rangle - \left\langle y\right\rangle^{2} \propto t^{\alpha_{y}}.
\end{gather}

\subsubsection{Kurtosis}

Kurtosis is defined as the standardized fourth moment of a distribution about its mean and is given by 

\begin{equation}
\mathrm{Kurt}[\vec{X}] = \frac{m_{4}}{\sigma^{4}} = \frac{\left\langle (\vec{X} - \vec{\mu})^{4}\right\rangle }{\left( \left\langle (\vec{X} - \vec{\mu})^{2}\right\rangle \right) ^{2}},
\end{equation}
where $m_{4}$ is the fourth moment, $\sigma$ the standard deviation, and $\vec{\mu}$ the mean. The kurtosis is a useful descriptor for a distribution's deviation from Gaussian. Gaussian distributions have a kurtosis of 3. Therefore, it is convenient to define $\gamma_{2}$, the \textit{excess kurtosis}, as $\gamma_{2} = \frac{m_{4}}{\sigma^{4}} - 3$.

% When $\gamma_{2} > 0$ the distribution is referred to as leptokurtic and is characterized by having heavier tails and higher peaks compared to normal distributions. Distributions with $\gamma_{2} < 0$ are referred to as platykurtic and are characterised as having lighter tails and flatter peaks than normal distributions \cite{DeCarlo1997}. Uniform distributions have a constant $\gamma_{2}$ equal to -1.2\cite{DeCarlo1997}. Furthermore a value of $\gamma_{2} = -2.0$ is the absolute minimum allowed value of excess kurtosis, and represents a distribution where all of the probability is located at the edges of the domain\cite{DeCarlo1997}. 

As our BBRs progress on their respective tracks, they produce time-evolving displacement distributions. In particular, we are interested in characterizing the differences between the evolving $x$- and $y$- components of the displacement distributions to understand the effect of constraints on the dynamics of BBRs. To this end, $\gamma_{2}$ provides a measure of the shape of the distributions and allows for easy comparison across a large parameter space of BBR designs. 

\section{\label{sec:level1}Results}
\subsection{\label{sec:level2}Sample trajectories and distributions}

\begin{figure}[h]
	\includegraphics[width=0.50\textwidth]{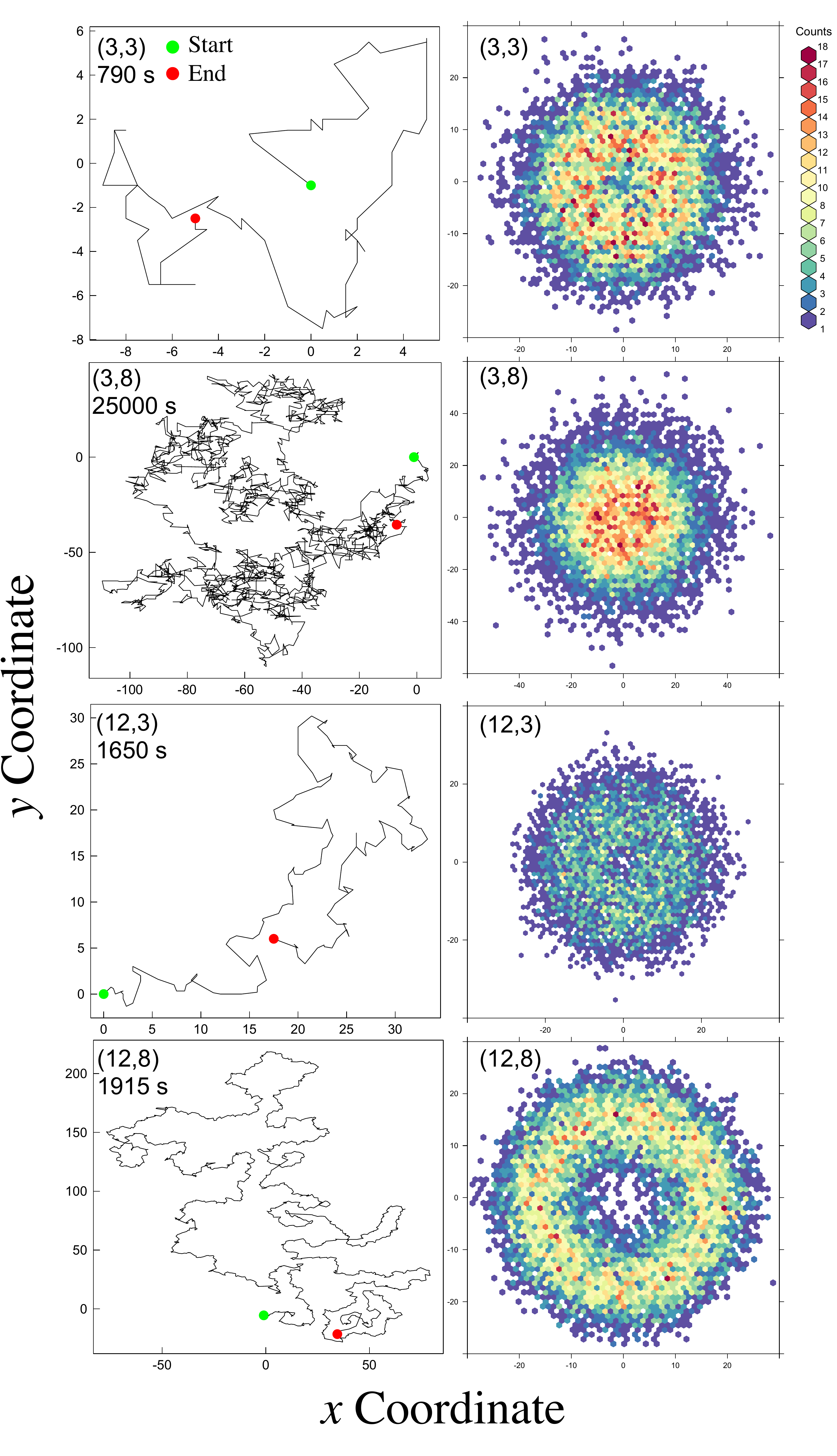}
	\caption{Sample trajectories and distributions of polyvalent BBRs on a two-dimensional track. Left column: sample trajectories. Green and red points indicate the starting and detachment positions for each respective trajectory; the time listed is the lifetime of that trajectory. (3,3), (12,3) and (12,8) BBRs are less likely to cross previously visited territories and exhibit more directional walks compared to (3,8) BBRs. Right column: snapshot of ensemble BBR position distributions, taken from \textbf{Movies S1-S4}. Color coding represents the number of independent BBRs at that location in time. All BBR designs, except for (3,8), develop ring-like distributions with decreased occupancy at the origin.} 
	\label{trajs}
\end{figure}

Single trajectories in two dimensions for (3,8), (3,3), (12,3) and (12,8) BBRs are presented in \textbf{Fig. \ref{trajs}} (left column). Our lowest span BBR systems, (12,3) and (3,3) BBRs, tend to become easily entrapped by their product wakes, leading to detachment. With increased span and decreased polyvalency, such as the (3,8) system, BBRs can reach over their previous trajectories into areas of fresh substrate.   

Snapshots of the ensemble behaviour of each BBR design in two dimensions are included in \textbf{Fig. \ref{trajs}} (right column), beside their respective sample trajectories. The (3,3), (12,3) and (12,8) systems all develop a low occupancy near their centre starting position, while the (3,8) system maintains the highest BBR occupancy around the origin. These ensemble snapshots are taken from \textbf{Movies S1-S4}, which present the full dynamic evolution of the ensemble behaviour.

\subsection{\label{sec:level2}Mean squared displacement}
 
\begin{figure}[h]
 	\includegraphics[width=0.5\textwidth]{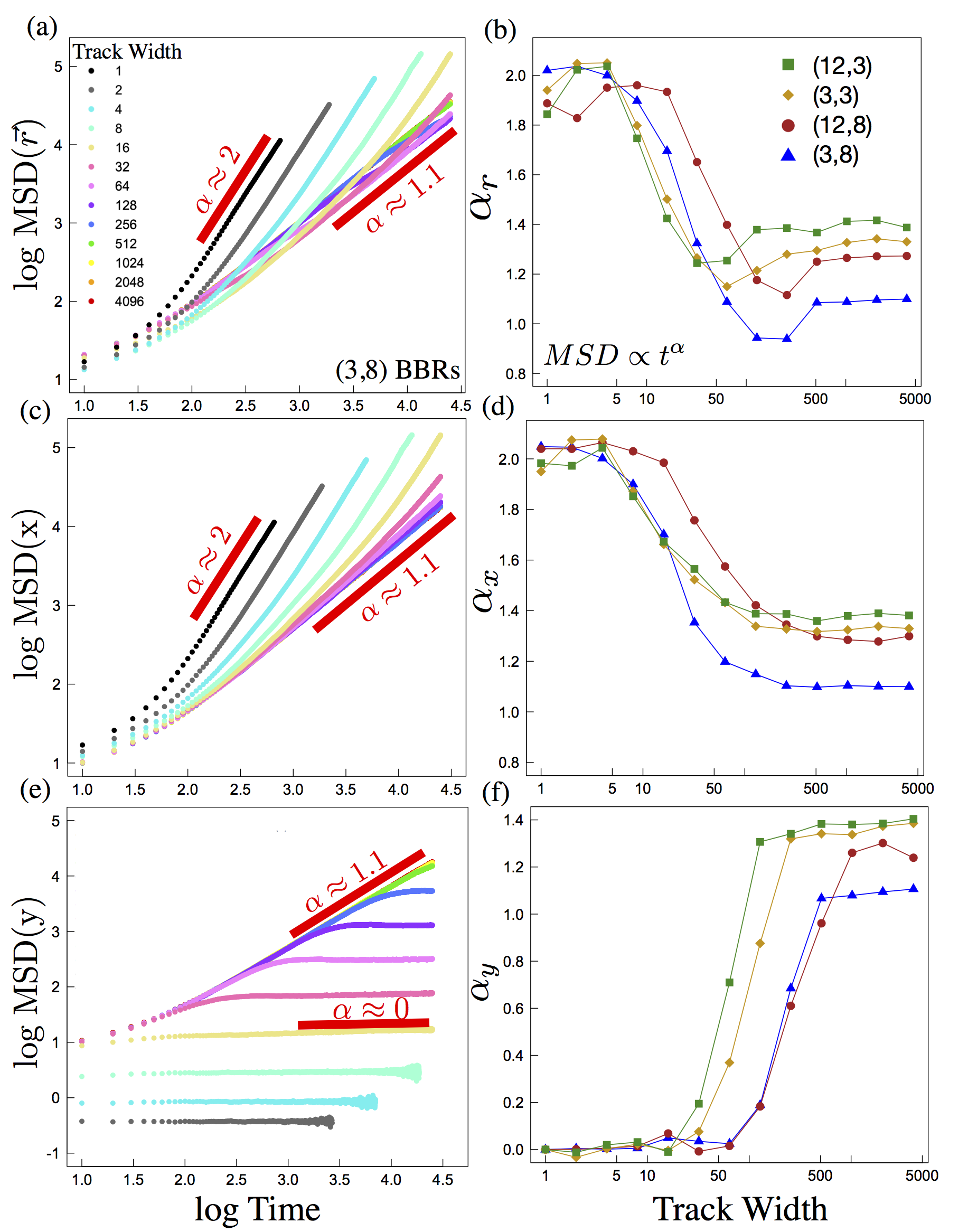}
 	\caption{a) log-log MSD($\vec{r}$)-time plots for (3,8) BBRs display the general trend of decreasing slope as a function of increasing track width. b) $\alpha_{r}$ for (3,8) BBRs reaches a minimum at track widths of 128-256 before subsequently increasing to a width-invariant plateau of $\alpha_{r} \approx 1.1$. A minimum $\alpha_{r}$ at intermediate widths is observed for all BBR designs examined in this work. c) MSD($x$) for (3,8) BBRs scales from ballistic to superdiffusive as the track width increases from one to two dimensions. d) $\alpha_{x}$ as a function of width for all BBRs displays a monotically decreasing trend. e) MSD($y$) for (3,8) evolves to $\alpha_{y} = 0$ for all BBRs that are constrained by track boundaries. f) For narrow tracks the BBRs move under confinement and display $\alpha_{y} \approx 0$. As track width increases such that the BBRs do not interact with the boundaries, $\alpha_{y}$ takes on the same values as $\alpha_{x}$ for all BBR designs.}
 	\label{MSDData}
 \end{figure}

 \textbf{Fig. \ref{MSDData}a} shows log-log plots of MSD($\vec{r}$) vs. time for (3,8) BBRs on all examined track widths. We report values of $\alpha$ in the long-time limit when $\dot{\alpha} \approx 0$. A one-dimensional track results in ratchets moving ballistically with $\alpha \approx 2$. As track width increases $\alpha$ begins to decrease. One would na\"{i}vely expect that as the width of the track increases, the constructs have increased probability to change direction, thus lowering $\alpha$. However, this transition does not occur monotonically. \textbf{Fig. \ref{MSDData}b} shows that we observe a minimum in $\alpha_{r}$ as a function of track width for all BBR designs.
 
 The non-monotonic behaviour of $\alpha_{r}$ as a function of track width prompted closer inspection of the MSD. The log-log plot of MSD($x$) (\textbf{Fig. \ref{MSDData}c}) depicts the expected MSD power law behaviour: as the track width increases $\alpha_{x}$ is found to decrease monotonically to a width-independent minimum (\textbf{Fig. \ref{MSDData}d}). By contrast, log-log MSD($y$)-time (\textbf{\textbf{Fig. \ref{MSDData}e}}) attains a slope of $\alpha_{y} \approx 0$ for narrow track widths in long-time limits (\textbf{Fig. \ref{MSDData}f}). 
 
 \textbf{Fig. \ref{MSDData}b} also shows that the large track width values of $\alpha_{r}$ depend on polyvalency and span. Short span and large polyvalency results in the highest $\alpha_{r}=1.4$, whereas the lowest $\alpha_{r}=1.1$ is found by lowering polyvalency and increasing the span. All of our examined BBR designs display similar MSD trends as a function of track width, as shown in \textbf{Fig. S1}.
 
\subsection{\label{sec:level2}Detachment curves}

\begin{figure}[h]
	\includegraphics[width=0.5\textwidth]{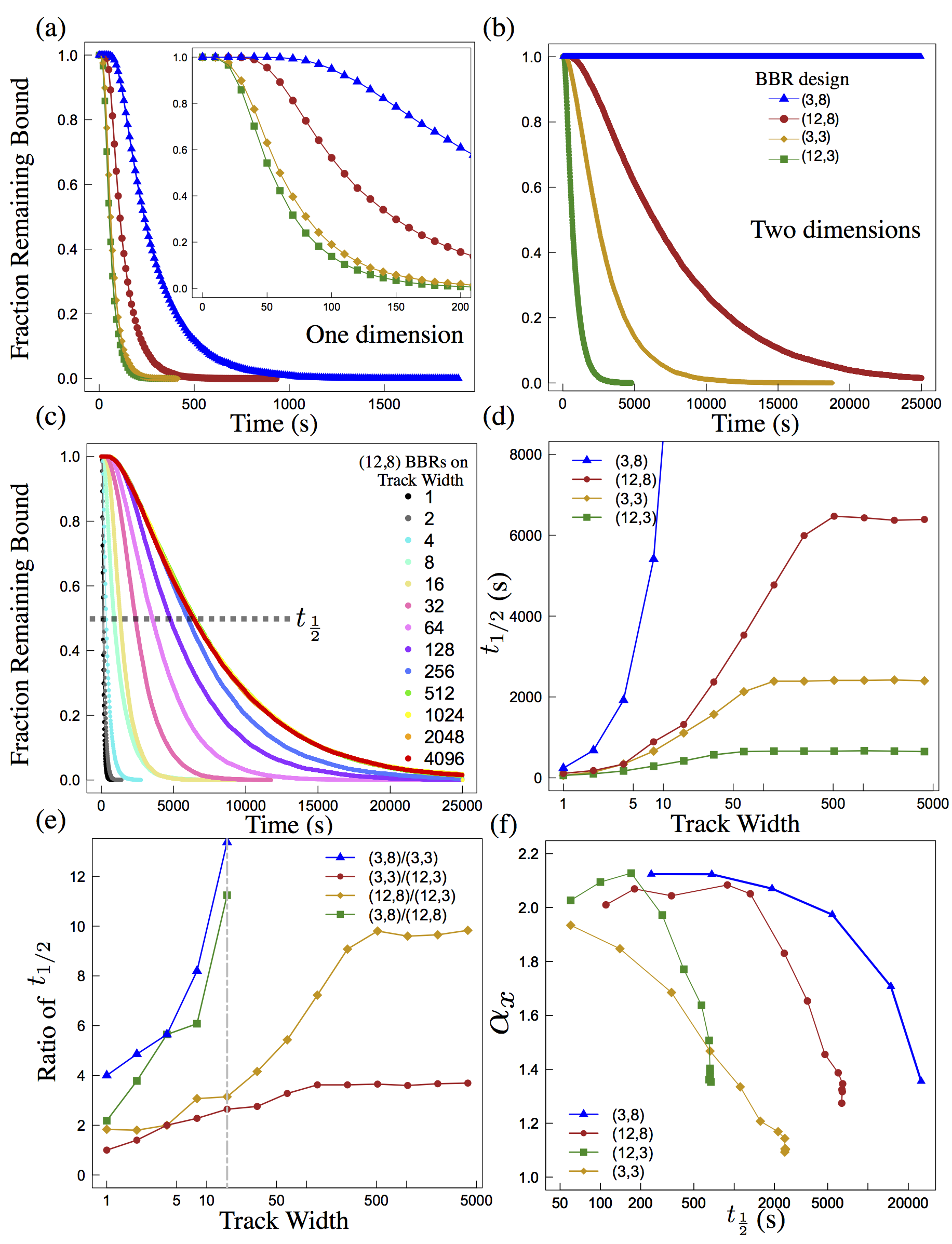}
	\caption{a) Fraction of BBRs remaining bound in one dimension for all BBR designs examined in this work. The inset shows the early time behaviour. b) Fraction of BBRs remaining bound in two dimensions for all BBR designs examined. For both 1D and 2D tracks (3,8) BBRs remain associated to the track for the longest times, whereas (12,3) BBRs associate to the track for the shortest times. c) Fraction of (12,8) BBRs remaining bound for all track widths. The detachment curves saturate past track widths of 256. d) $t_{1/2}$ as a function of track width for all BBR designs. e) The ratio of $t_{1/2}$ values for various BBR designs. f) $\alpha_{x}$ values from \textbf{Fig. \ref{MSDData}d} plotted against their respective $t_{1/2}$ values.}
	\label{tHalfAlpha}
\end{figure}

In our simulations we do not allow for re-attachment once all of the legs of the BBR have detached. Thus, it is useful to investigate how track association times vary with polyvalency and span. \textbf{Fig. \ref{tHalfAlpha}ab} display the fraction of BBRs remaining bound for all examined BBR designs in the 1D and 2D track limits. We find that (3,8) BBRs remain associated to the track for the longest times, whereas (12,3) BBRs detach the fastest. In \textbf{Fig. \ref{tHalfAlpha}c} we show that track association time increases monotonically as a function of increasing track width for (12,8) BBRs. The detachment curves saturate and overlap for track widths larger than 256. All BBR designs display a similar trend, and can be viewed in \textbf{Fig. S2}. 

For all detachment curves we define $t_{1/2}$ as the time at which 50\% of the BBRs have detached from the track. For (3,8) BBRs we observed negligible detachment on tracks of width greater than 32, therefore we cannot report $t_{1/2}$ values for wider tracks. \textbf{Fig. \ref{tHalfAlpha}d} depicts $t_{1/2}$ as a function of track width for each BBR design. We can see that both polyvalency and span have large effects on the observed $t_{1/2}$. Across all track widths (12,3) BBRs consistently detach faster than all other BBR designs. For (12,3) ratchets, increasing track width from one to two dimensions results in an increase of $t_{1/2}$ by a factor of 10 (\textbf{Fig. \ref{tHalfAlpha}d}). A similar comparison of (12,8) BBRs yields a factor of 60 increase in track association time.  Therefore, in the limit of large polyvalency and short span we see less of a gain in track attachment time by increasing the track width than for larger span. 

 We next compared the different designs directly by taking ratios of the $t_{1/2}$ trends as shown in \textbf{Fig. \ref{tHalfAlpha}e}. In the limit of low polyvalency, increasing span from 3 to 8 has a profound effect on increasing track association time. Similarly, in the limit of high span, decreasing polyvalency from 12 to 3 profoundly increases track association time. However, if the span is kept constant at 3, the decrease in polyvalency from 12 to 3 has little impact on track assocation time across all track widths. Therefore, the improvement in track association time gained from decreasing polyvalency is only realized for the BBR systems with large span.
 
In \textbf{Fig. \ref{tHalfAlpha}f} we plot $\alpha_{x}$ against $t_{1/2}$. For all BBR systems $\alpha_{x}$ tends to decrease with increasing $t_{1/2}$ values.
 
\subsection{\label{sec:level2}Excess kurtosis}

\begin{figure}[h]
	\includegraphics[width=0.5\textwidth]{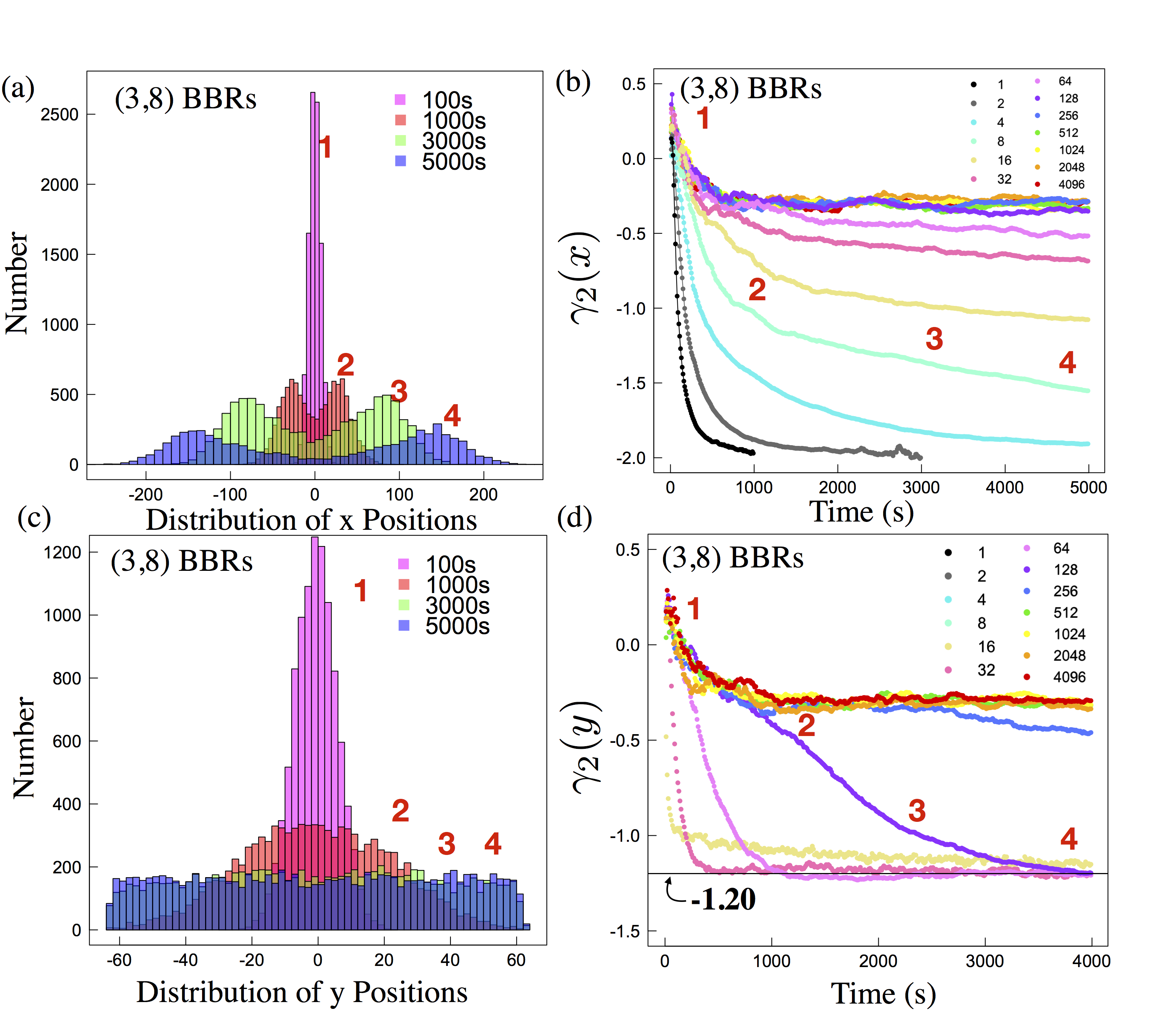}
	\caption{a) The evolving $x$-position distribution for (3,8) BBRs on a track width of 8. The numbers 1-4 represent the distribution at specific times. Two modes develop that move symmetrically in opposing directions. b) Excess kurtosis,$\gamma_{2}$, for $x$-displacement distributions for (3,8) BBRs on all track widths. On narrow tracks $\gamma_{2}(x)$ quickly approaches -2.0. As track width increases the excess kurtosis settles to $\gamma_{2}(x) =$ -0.25. The numbers 1-4 correspond to the histograms from a). c) The evolving $y$-position distribution for (3,8) BBRs on a track of width 128. d) For large track widths $\gamma_{2}(y)$ limits to -0.25, the same as $\gamma_{2}(x)$. Under confinement $\gamma_{2}(y)$ settles to -1.2, the value for a uniform distribution, as illustrated by the numbers 1-4 from part c).} 
	\label{allKurtosis}
\end{figure}

The ensemble-average displacement of the ratchets away from the origin results in evolving displacement distributions, as shown in \textbf{Movies} \textbf{S5} and \textbf{S6}. All of the BBRs are initialized at the centre of the track, therefore the initial distribution is peaked at the origin at $t = 0$ s. As BBRs progress along the track, their cleavage of track sites limits options for turning back. In one dimension, as the BBRs randomly break the track symmetry a bimodal distribution develops whose modes propagate in opposite directions. \textbf{Fig. \ref{allKurtosis}a} illustrates the typical development and separation of the two modes on a narrow track width of 8 for (3,8) BBRs. From \textbf{Movie S5}, qualitatively, one can see that the modes are both separating and dispersing with time. \textbf{Fig. \ref{allKurtosis}b} shows the results of computing the excess kurtosis, $\gamma_{2}(x)$, for these (3,8) BBRs on all track widths. In all cases $\gamma_{2}(x)$ initializes slightly higher than the Gaussian value of 0 as the distribution is initially peaked sharply around the origin. On narrow tracks $\gamma_{2}(x)$ rapidly reduces to the -2.0 limit, indicating a distribution whose probability is located at the edges of its domain. For wider tracks, $\gamma_{2}(x)$ reduces to -0.25. Similar behaviour is seen for the other BBR designs, as shown in \textbf{Fig. S3}. 

We next look at the correponding evolution of position distributions for the $y$ coordinates. The $y$-position distribution for (3,8) BBRs on a track width of 128 evolves into a uniform distribution across the accessible domain of lattice sites. To understand how this lateral shape of the distribution changes upon interacting with the boundary we compute $\gamma_{2}(y)$ (\textbf{Fig. \ref{allKurtosis}d}). For wide tracks, where the BBRs do not reach the boundaries, we see that $\gamma_{2}(y)$ approaches -0.25, the same as found for $\gamma_{2}(x)$. However, for (3,8) BBRs on a track width of 128, which begin to approach the track boundary around $\sim 1000s$, we see that $\gamma_{2}(y)$ decreases to -1.2, the value for a uniform distribution \cite{DeCarlo1997}. For these (3,8) BBRs $\gamma_{2}(x)$ remains constant at -0.25 once they have reached the y-boundaries. The shape of the $x$-displacement distribution is therefore not affected by interactions with the track boundary, while the $y$-displacement distribution clearly is. Similar behaviour is seen for the other BBRs (\textbf{Fig. S4}).

\subsection{\label{sec:level2}Bimodal Gaussian model}

%phenomalogical model that captures the basic behaviour of the evolution of BBR displacement distributions. 

To further our understanding of the time dependence of $\gamma_{2}(x)$ we analytically derived the excess kurtosis for a probability density function comprised of two Gaussian distributions with equal variance and equal but opposite means, as shown in eq. 6.
\begin{equation}
P(x|\mu, \sigma) = \frac{1}{2\sqrt{2\pi \sigma^{2}}}e^{-\frac{(x-\mu)^{2}}{2\sigma^{2}}} + 
\frac{1}{2\sqrt{2\pi \sigma^{2}}}e^{-\frac{(x+\mu)^{2}}{2\sigma^{2}}}
\end{equation}
We computed the 2nd and 4th moments of $P(x)$ to determine the excess kurtosis (see supporting information). We find $\gamma_{2}$ to be given by
\begin{equation}
 \gamma_{2} = \frac{2\mu^{4} + 12\mu^{2}\sigma^{2} + 6\sigma^{4}}{2(\mu^{2} + \sigma^{2})^{2}} - 3.
\end{equation} 
The MSD of this distribution is
\begin{equation}
 \mathrm{MSD}
  = \left\langle(x-\mu)^{2}\right\rangle = \mu^{2} + \sigma^{2}.
\end{equation} 

We examine four cases in our analytical model:
\begin{align*}
\mathrm{a)}& \; \mathrm{Constant\; separation,\; Brownian \; dispersion}: \\ & \mu = \mu_{0}; \; \sigma^{2} = Dt \\
\mathrm{b)}& \; \mathrm{Linear \; separation,\; constant\; dispersion}: \\ & \mu = bt; \; \sigma^{2} = D_{0}\\
\mathrm{c)}& \; \mathrm{Linear \; separation,\; Brownian \; dispersion}:\\ & \mu = bt; \; \sigma^{2} = Dt\\
\mathrm{d)}& \; \mathrm{Linear \; separation \; \& \; dispersion}: \\ 
 &\mu = bt; \; \sigma = D_{1}t
\end{align*}

 For each of the above cases, eq. 8 can be expressed as a quadratic equation with different coefficients,  
 \begin{equation}
  \mathrm{MSD} = a_{0} + a_{1}t + a_{2}t^{2}.
 \end{equation} 
  Tuning the linear and quadratic coefficients can be used to tailor the power-law scaling $\alpha$. For example, if $b = 0$ and $D > 0$ we achieve conventional diffusion ($\alpha = 1$), whereas if $D = 0$ and $b > 0$ we get ballistic motion ($\alpha = 2$). If both $b>0$ and $D>0$ the type of diffusion depends on the ratio $\frac{D}{b}$, which dictates the timescales of interest.  
  
\begin{figure}[h]
	\includegraphics[width=0.5\textwidth]{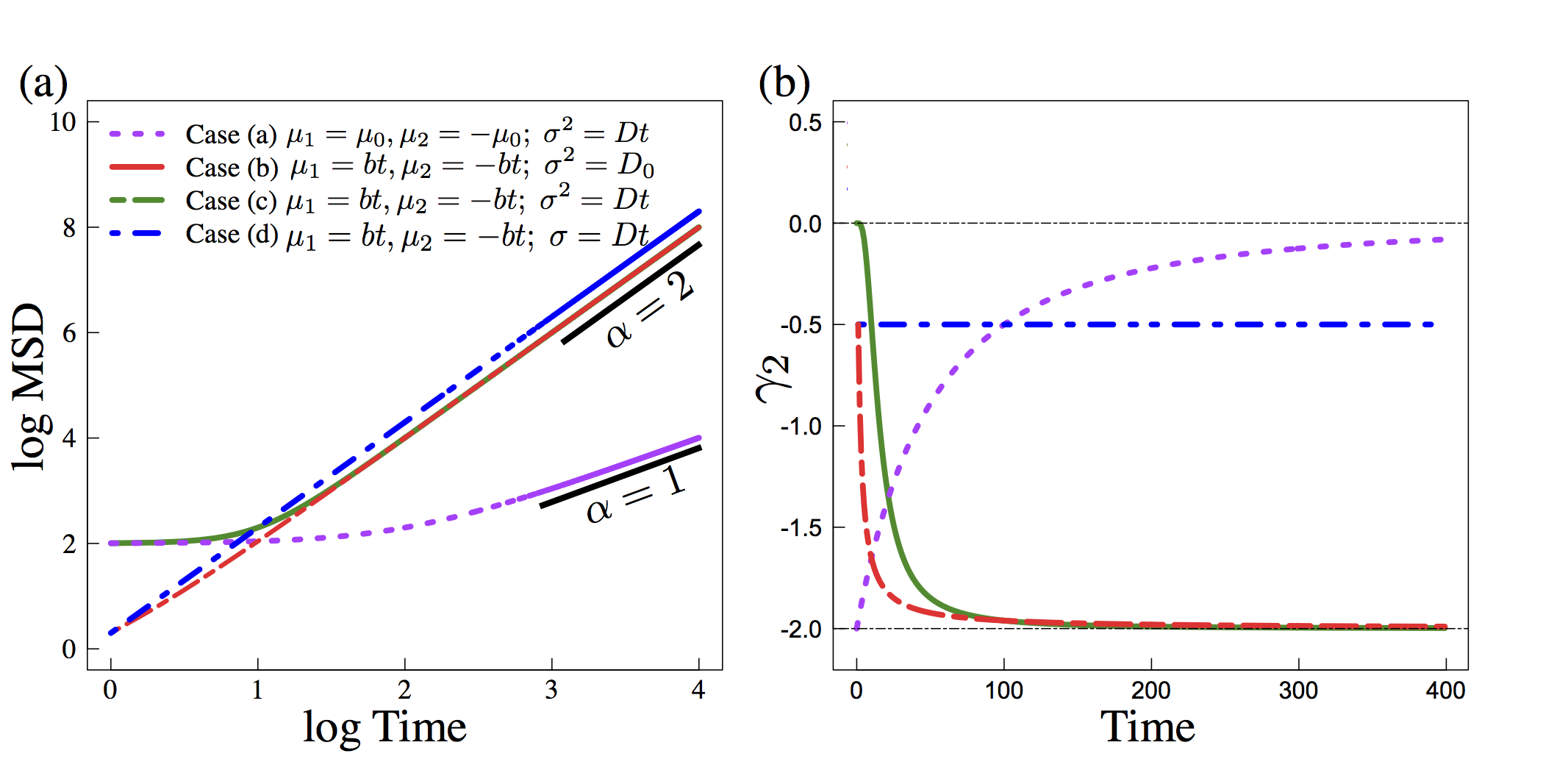}
	\caption{Results of bimodal Gaussian model. a) In the limit of Brownian dispersion (case a) we recover conventional diffusion at long times. For all other cases we find the long-time limit of $\alpha$ to represent ballistic motion. b) With increasing $\sigma(t)$ and constant $\mu$ (case a), $\gamma_{2}$ reaches the Gaussian limit of $\gamma_{2} = 0$. In cases b and c, which describe distributions that separate faster than they disperse, $\gamma_{2}$ decreases to -2.0. When $\sigma$ and $\mu$ increase at the same rate $\gamma_{2}$ remains constant (case d).} 
	\label{MSDKurt}
\end{figure}

   In \textbf{Fig. \ref{MSDKurt}} we take $D = D_{1} = b = 1$, $\mu_{0} = D_{0} = 10$, and compute MSD (\textbf{Fig. \ref{MSDKurt}a}) and $\gamma_{2}$ (\textbf{Fig. \ref{MSDKurt}b}). For case a, we find the system exhibits subdiffusive motion for timescales up to $10^{3}$ seconds (\textbf{Fig. \ref{MSDKurt}a}). We do not see conventional diffusion unless we compute $\alpha$ at longer times. Similarly, for case c, subdiffusive behaviour is observed at short timescales followed by a crossover to ballistic motion at longer times. 
  
  The excess kurtosis also varies by case. For case a, where the mean of the modes is fixed to $\pm \mu_{0}$ with Brownian dispersion, $\gamma_{2}$ increases from -2.0 to the Gaussian limit of 0. With cases b and c, where the modes are separating faster than they are dispersing, $\gamma_{2}$ reduces from 0 to -2.0. Lastly for case d, where $\mu(t)$ and $\sigma(t)$ are both linearly increasing at the same rate, $\gamma_{2}$ takes on a constant value of -0.5.
   
\subsection{\label{sec:level2}Substrate digestion rates}

\begin{figure}[h]
	\includegraphics[width=0.5\textwidth]{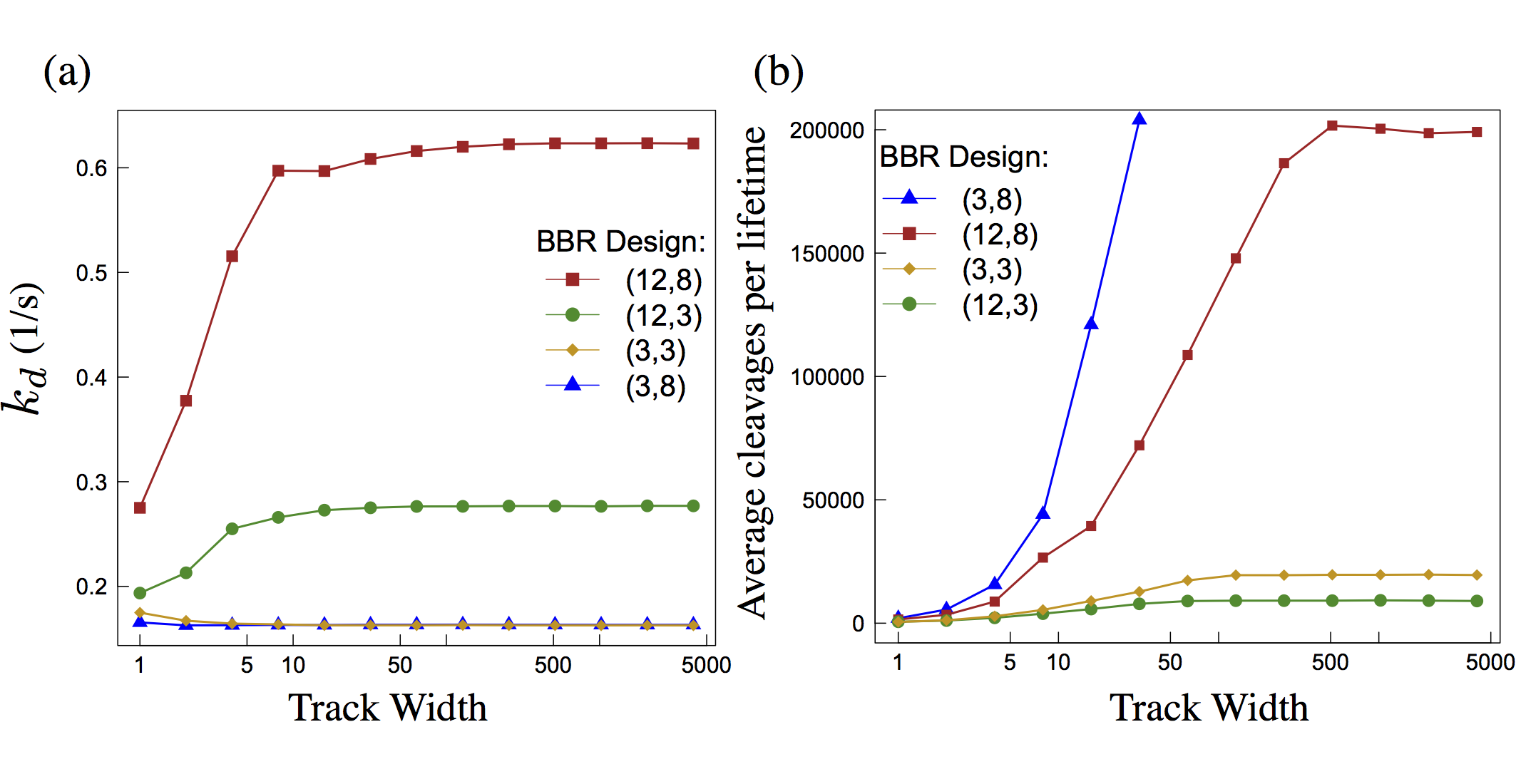}
	\caption{a) Digestion rate, $k_{d}$, for each BBR design across all track widths. For all BBR designs substrate digestion rates are found to be width-independent for tracks larger than a width of $\sim$8. b) Average number of substrate sites digested before detachment (starvation) for all BBRs across all examined track widths. For (3,8) BBRs, average cleavages are reported for tracks only up to width 32, as most (3,8) BBRs remained bound on larger track widths.} 
	\label{cleavages}
\end{figure}

Those who study starved random walks are often concerned with the number of food items the walker consumes before starvation \cite{Benichou2014}. In our model starvation can be defined as the walker having no accessible substrate `food' sites within its span. While we have already characterized the total time associated to the tracks, here we characterise track digestion rates and total successful cleavages before starvation (detachment). We define the substrate digestion rate, $k_{d}$, as the average number of cleavages observed for each BBR design per second. $k_{d}$ is distinct from $k_{eff}$ which is known \textit{a priori} and used to simulate cleavage kinetics in the Gillespie algorithm.\\
\indent In \textbf{Fig. \ref{cleavages}a} we report $k_{d}$ for all examined BBRs across all track widths. Not surprisingly, we find polyvalency to be the dominating factor for increasing the digestion rate, where both (12,8) and (12,3) BBRs have the highest $k_{d}$. Interestingly, cleavage rates for the 12-legged BBRs increase with track width to a constant value, whereas 3-legged BBRs experience a slight decrease in their cleavage rates as the width of the track is increased. When we examine the average number of cleavages before detachment, the inherent track association time, characterised by $t_{1/2}$, is the dominating system parameter. \textbf{Fig. \ref{cleavages}b} displays average cleavage events vs. track width, which scales similary to $t_{1/2}$ (\textbf{Fig. \ref{tHalfAlpha}d}). 

\section{\label{sec:level1}Discussion}
\subsection{\label{sec:level2}Mean squared displacement}

The MSD holds information on the BBR's ability to move directionally, a key design criterion for synthetic molecular motors. MSD scales as $t^{\alpha}$, where $\alpha$ characterizes the type of anamolous diffusion inherent to the system \cite{Havlin2002}. For values of $\alpha$ ranging from $0 < \alpha < 1$ the motion is subdiffusive, which describes characteristic motion slower than that of conventional diffusion. A value of $\alpha = 1$ describes a system that undergoes conventional diffusion. Systems with $1 < \alpha < 2$ are superdiffusive, a property of systems that undergo active transport. When $\alpha = 2$ the system exhibits linear (ballistic) motion and is ideal for molecular transport systems. 

 We find the long-time $\alpha$ values to be highly dependent on track width, span, and polyvalency. We expected $\alpha$ to maximize on narrow tracks for each BBR design, as the effect of confinement would promote linearly directed motion. With reference to \textbf{Fig. \ref{MSDData}bd}, we find that $\alpha$ is maximum for narrow tracks and persists with near ballistic values for tracks of width larger than 1 lattice site. For (12,8) BBRs, $\alpha_{x}$ remains constant for track widths up to 16, whereas for (3,8) BBRs $\alpha_{x}$ begins to decrease at a track width of 4 (half the span for this system) (\textbf{Fig. \ref{MSDData}d}). Therefore, both polyvalency and span play a role in maintaining optimally ballistic motion as the effects of confinement are relaxed.\\
\indent At early times all examined BBRs display subdiffusive behaviour at timescales proportional to $\frac{1}{k_{eff}}$, as was also reported by Olah et al. \cite{Olah2013}. We observe initially subdiffusive behaviour for all track widths. This makes intuitive sense as each run is initialized with one leg associated to the track. Unbound legs then need to bind and cleave in order to translocate the global constraint, which by design occurs on a timescale of $\frac{1}{k_{eff}}$ regardless of the effects of confinement. \\
 \indent In \textbf{Fig. \ref{MSDData}f} we report $\alpha_{y}$ values from fitting to the late-time MSD trends when the BBRs have had sufficient time to reach the boundaries. As the track width increases, the characteristic time for $\alpha_{y}$ to transition to 0 increases. When $\alpha_{y} = 0$, $\mathrm{MSD}(y)$ takes on the variance of a uniform distribution whose domain is defined by the track boundaries. \\
 \indent The surprising result in \textbf{Fig. \ref{MSDData}b}, where $\alpha_{r}$ develops a minimum as a function of width, can then be explained by interactions with the boundary, which impose a sub-diffusive characteristic on $\alpha_{r}$. In contrast, when the BBRs are not constrained by the track boundaries we find that $\alpha_{r} = \alpha_{x} = \alpha_{y}$.\\
 \indent For effectively two-dimensional tracks, larger span and lower polyvalency, such as the (3,8) BBRs, results in a lower $\alpha_{r}$ of 1.1. The evolution of the ensemble of (3,8) BBRs, as shown in \textbf{Movie S2}, also displays a permanent high occupancy around the starting position.  Conversely, higher polyvalency and shorter span, such as the (12,3) BBRs, results in a greater $\alpha_{r}$ of 1.4. For the (12,3) system, the evolution of the ensemble distribution (\textbf{Movie S3}) also displays low occupancy around the starting position, leading to a ring-like structure in the two-dimensional distribution. The emergence of the ring structure indicates that the ensemble exhibits radially directed motion, away from the starting position.
 
 Increased polyvalency therefore leads to the most superdiffusive walk in two dimensions. Why is this so?  In our Gillespie model, the rate of binding to substrate, $k_{on}$, is $\sim$400 times higher than the effective rate of cleavage, $k_{eff}$. Therefore, all unbound legs will preferentially bind to locally available substrate sites. Each unbound leg acquires a transition rate, $k_{on}$, for each of the available $N$ substrate sites. For (12,3) BBRs this means that all legs will preferentially saturate the track. The increased number of track-associated legs means that the product wake produced by the BBR is also denser. By contrast, (3,8) BBRs can access larger regions of the track with each step, and with a polyvalency of only 3 their product wake is expected to be sparse. By this reasoning, the (12,3) BBRs are expected to have higher $\alpha_{r}$ values than the (3,8) BBRs because it is harder for the global constraint to change direction; multiple legs need to coordinate to move the global constraint towards a new direction, leading to a higher value of $\alpha_{r}$. 
 
  Substrate track sites that have previously been visited, and subsequently turned to product-sites, cannot be revisited. However, the global constraint can still visit its previous locations and cross over its path because the legs can bind beyond their nearest neighbours. The more times the global constraint revisits a location the less likely it will be to return because the local region becomes further depleted of substrate. Therefore, despite the ideal burnt-bridges behaviour of each leg, our BBRs do not scale as a strict self-avoiding walk. There may be merit in the application of models for \textit{weakly} self-avoiding walks to polyvalent BBRs \cite{Madras1996}. The BBR system may also bear relevance to foragers eating a subset of food per site \cite{Redner2018}. 
   
\subsection{\label{sec:level2}Track dissociation and its effect on $\alpha$}

	\indent The width of the track has a strong effect on both the observed $\alpha$ values and the track association time, as shown in \textbf{Fig. \ref{MSDData}} and \textbf{Fig. \ref{tHalfAlpha}}. As stated $\alpha_{x}$ does not begin to decrease until a track width of 16 for (12,8) BBRs (\textbf{Fig. \ref{MSDData}d}). However, their track association time increases dramatically from a width of 1 to a width of 16 (\textbf{Fig. \ref{tHalfAlpha}c}). To understand the relationship between $\alpha_{x}$ and $t_{1/2}$ we first looked to see if they are correlated. In \textbf{Fig. \ref{tHalfAlpha}f} we see that $\alpha_{x}$ is relatively constant and independent of $t_{1/2}$ for these BBRs on tracks of width 1-16. $t_{1/2} \approx 100$ s on a track of width 1, whereas on a track of width 16 $t_{1/2} \approx 2000$ s. By increasing the width of the track to twice that of the BBR span, the (12,8) system maintains ballistic behaviour while gaining more than an order-of-magnitude increase in track attachment time. A further increase in track width from 16 to 4096 results in an increase of $t_{1/2}$ by a factor of 2.5, but a decrease in $\alpha_{x}$ from $\sim 2.0$ to $\sim 1.3$. We therefore conclude that if one wishes to increase both directionality and track association time, designing tracks of a width proportional to the span of the ratchet is optimal. When one further increases the track width, $\alpha$ begins to decrease. \\
	\indent Both decreasing polyvalency and increasing span result in increased track attachment time. Of the two design parameters, which has the strongest impact on maintaining track assocation? To illustrate the effects of polyvalency on track association we can compare (3,8) and (12,8) BBRs on tracks of width 1 and 8. On a track width of 1, we find $t_{1/2}$ to be 240 and 110 seconds for (3,8) and (12,8) BBRs, respectively (\textbf{Fig. \ref{tHalfAlpha}d}). However, the effect of polyvalency on track association highly depends on the span, which we map out in \textbf{Fig. \ref{tHalfAlpha}e}. For example, increasing polyvalency of span-3 BBRs results in at most a factor of 2 increase in $t_{1/2}$, even on two-dimensional tracks. However, for span-8 BBRs, $t_{1/2}$ increases by a factor of 2 for one-dimensional tracks, but by a factor of 10 for two-dimensional tracks when the number of legs is increased from 3 to 12. Altering span also shows similar trends where the gain in $t_{1/2}$ highly depends on the polyvalency.\\
	\indent It may seem counterintuitive that an increase in polyvalency leads to a decrease in $t_{1/2}$ given previous work with molecular spiders \cite{Pei2006,Samii2011}. However, in molecular spider systems the walkers have a rate of product binding, $k_{onP}$, typically taken to be the same as substrate binding, $k_{onS}$ \cite{Olah2013,Samii2010,Samii2011}. Therefore, when the walker digests all local substrate sites it can search through local product sites for areas of fresh substrate. In such a system, increased polyvalency means more options for product site coupling and subsequently decreased probability of detachment. By contrast, our system is a ideal BBR where $k_{onP} = 0$, and we find increased polyvalency leads to a decrease in track attachment time. 
	  
 \subsection{\label{sec:level2}Excess Kurtosis}
 
The shape of the BBR position distribution has a time dependence. Kurtosis is a convenient measure to compare BBR position distributions across our parameter space of width, polyvalency, and span. On narrow tracks the position distribution for (3,8) BBRs immediately develops into two modes that move in opposite directions. This is reflected as a monotonic decrease in $\gamma_{2}(x)$ to -2.0 (\textbf{Fig. \ref{allKurtosis}b}), which indicates a distribution with probability isolated to the edges of the domain. The splitting of the position distribution into two oppositely moving modes is consistent with the ballistic behaviour characterized by $\alpha$. 
 
 Having seen that the distributions formed two oppositely moving modes inspired us to analytically derive kurtosis for a PDF described by two Gaussians, as shown in eq. 6. We analytically derived $\gamma_{2}$ (eq. 7) in one dimension and explored the effects of dispersion and mode separation, as shown in \textbf{Fig. \ref{MSDKurt}}. Increasing dispersion, while maintaining a constant separation of modes, leads to $\gamma_{2}$ monotonically increasing towards the Gaussian value of 0, indicating that the two independent modes are completely overlapping. Conversely, a linear increase in mode separation with constant dispersion results in $\gamma_{2}$ monotonically decreasing to -2.0. Furthermore, dispersion and mode separation have compensatory effects on $\gamma_{2}$, such that if they are increasing at similar rates $\gamma_{2}$ remains constant. From our phenomenological Gaussian model we are able to reproduce all of the BBR $\gamma_{2}$ results by modifying $\sigma$ and $\mu$. 
 
 Our analytical model offers insights into the BBR behaviour and allows us to understand the effects of polyvalency, span and width on the shape of the position distributions. For example, $\gamma_{2}(x)$ for (3,8) BBRs on wide tracks reaches a time-invariant value of -0.25 (\textbf{Fig. \ref{allKurtosis}b}). From our bimodal Gaussian model, this suggests that the dispersion and separation of the modes are equal. We verify this for the (3,8) system where we compute $\sigma(t)/\mu(t)$ (\textbf{Fig. S5}). The ratio of $\sigma(t)/\mu(t)$ is known as the coefficient of variation. In our analytical system the coefficient of variation is equal to $\frac{D}{b}$ where $D$ and $b$ are equivalent to diffusion and drift coefficients, respectively. The ratio $\frac{D}{b}$ is a ratio of diffusion and mobility, and has been used to characterize the ability of Brownian ratchets to achieve directional motion under external fields \cite{VanOudenaarden1999}.  
 
 %The behaviour of $\gamma_{2}(x)$ highly depends on the BBR parameter space. (12,8) BBRs, which have both maximum span and polyvalency, shows different $\gamma_{2}(x)$ behaviour on wide tracks (\textbf{Fig. S3}). In this system, $\gamma_{2}(x)$ first develops into a peak across the first 500 s. Following this initial behaviour, $\gamma_{2}(x)$ then proceeds to increase slowly over thousands of seconds. Eventually this slow increase turns over to a constant value. Increasing both span and polyvalency therefore has the ability to impose long term effects on BBR dispersion. The mechanism for this effect is currently unknown to us.
  
 On wide tracks, for all BBR designs, $\gamma_{2}(y)$ takes on the same values as $\gamma_{2}(x)$ (\textbf{Figs. S3,S4}). As the BBRs are constrained by the boundaries, the position distributions evolve into uniform distributions across the width ($\gamma_{2}(y) = -1.2$), consistent with our finding that the variance given by MSD($y$) approaches that of a uniform distribution. 
 
  \subsection{\label{sec:level2}Substrate digestion rates}
 
Across all track widths we find that BBRs with larger span and polyvalency have the highest substrate digestion rate ($k_{d}$), as shown in \textbf{Fig. \ref{cleavages}a}. The ratio of the binding and effective cleavage rates used in the Gillespie model, $k_{on}$ and $k_{eff}$, respectively, is $\sim$400. This means that an unbound leg that can access a fresh substrate site is going to be 400 times more likely to bind to any one site than a bound leg is to cleave and release. Thus, all legs are likely to be bound to available substrate sites. (12,8) BBRs have access to more local substrate than the other BBR designs given that they have the largest polyvalency and the longest reach (span), therefore (12,8) BBRs are expected to cleave the most per unit time. (12,3) BBRs are `second best' to (12,8) BBRs with regards to the substrate digestion rate. All 12 legs can saturate to the track, but this design suffers from a shorter span leading to an inability of (12,3) BBRs to reach distant patches of fresh substrate, as compared to the (12,8) system.

It was surprising to see that (3,3) and (3,8) BBRs experienced a slight decrease in average substrate digestion rates as a function of increasing track width (\textbf{Fig. \ref{cleavages}a}). The decline is slight: from  0.18 $s^{-1}$ in one dimension, to 0.16 $s^{-1}$ in two dimensions for (3,3) BBRs. As the width is increased one would naively think that more substrate should be available to the BBR for any given combination of bound legs, therefore resulting in an increase digestion rate. This was not found for three-legged BBRs. Furthermore, for any given track width, the average digestion rates for (3,3) and (3,8) BBRs are within 5\%, suggesting that in the limit of low polyvalency, span plays no significant role in altering the substrate digestion rate. 

  Our hypothesis is that in one dimension, when the span-3 walkers move into their product wake, they quickly detach as there is little opportunity to turn around towards fresh substrate. However, as the track width is increased the walkers have more opportunity to rescue themselves from a substrate-barren environment. We speculate that on average, the (3,8) and (3,3) walkers experience more substrate-barren terrain in wider track widths, which leads to a lower average substrate digestion rate as they spend more time rescuing themselves from locally depleted regions. 
  
 The BBRs studied in this work can be considered as polyvalent depletion-controlled foragers \cite{Benichou2014, Redner2016}.  In \textbf{Fig. \ref{cleavages}b} we plot the average number of substrate sites digested by each BBR on each track width. As a function of track width, substrate digestion per lifetime scales similarly as $t_{1/2}$ (\textbf{Fig. \ref{tHalfAlpha}d}). Despite (3,8) BBRs having the lowest $k_{d}$ across all track widths, their greater track association time leads to them having more time to digest substrate, resulting in the most substrate cleaved prior to detachment.
  
 \section{\label{sec:level1}Conclusions}
 
 The design and implementation of synthetic machinery has shown great promise towards the control of motion at the nanoscale. In particular, synthetic analogues of biological molecular motors that implement a BBR mechanism have made great progress. Our goal in this work was to explore the effects of confinement on BBR performance and to provide design insights for \textit{de novo} BBR motors. Our results offer guidelines for researchers to follow when thinking about optimizing particular BBR characteristics. To fabricate a superdiffusive BBR in two dimensions, one should increase polyvalency and decrease span, as has been done in some systems \cite{Vecchiarelli2014a,Yehl2015}. Increasing span and decreasing polyvalency, in contrast, results in large increases to track attachment time but decreased directionality. Furthermore, we found that narrow tracks result in ballistic dynamics, as well as an order-of-magnitude increase in track attachment time compared to a one-dimensional track. Lastly, we found that increasing polyvalency results in an increased rate of substrate digestion, however, the total average track association time is the dominant factor that dictates total cleavage events before detachment. Through exploring the dimensionality-dependent crossover in motility of polyvalent BBRs, we have found these systems to exhibit rich dynamics. We hope these results provide useful insight towards the design of \textit{de novo} BBR systems.
 
  \section{\label{sec:level1}Acknowledgements}

This work was funded by the Natural Sciences and Engineering Research Council of Canada (NSERC), through a Discovery Grant to NRF. Computational resources were provided by Compute Canada.

\bibliographystyle{ieeetr}
\bibliography{References.bib}

\end{document}

\maketitle % Insert the title, author and date

\begin{center}
%\begin{tabular}{l r}
\begin{tabular}{c}
Chapin S. Korosec (ckorosec@sfu.ca) \\
Martin J. Zuckermann (mjzheb@gmail.com)\\
Nancy R. Forde (nforde@sfu.ca) \\
\end{tabular}

%\author{Corresponding authors:\\ ckorosec@sfu.ca \\ nforde@sfu.ca} % Author
\end{center}

\thispagestyle{empty}
\newpage

% If you wish to include an abstract, uncomment the lines below
% \newpage

% \tableofcontents

\newpage

\section{Kurtosis of a mixture of two Gaussian distributions}

On narrow tracks the BBR displacement distributions produce two modes that travel in the $\pm \hat{x}$ directions. To understand how the kurtosis is expected to behave we derive $\gamma_{2}$ for a probability density function described by two Gaussian modes centred at $\mu_{1}$ and $\mu_{2}$ with standard deviations $\sigma_{1}$ and $\sigma_{2}$, 
\begin{equation}
P(x| \mu_{1}, \mu_{2}, \sigma_{1}, \sigma_{2}) = \frac{1}{2\sqrt{2\pi \sigma_{1}^{2}}}e^{-\frac{(x-\mu_{1})^{2}}{2\sigma_{1}^{2}}} + 
\frac{1}{2\sqrt{2\pi \sigma_{2}^{2}}}e^{-\frac{(x-\mu_{2})^{2}}{2\sigma_{2}^{2}}}.
\end{equation}
$P(x)$ from eq. 1 can be used to compute the kth moment about the mean according to
\begin{equation}
m_{k} = \int_{-\infty}^{\infty}(x-\mu)^{k}P(x)dx.
\end{equation}
Kurtosis is defined as the standardized 4th moment. Since a Gaussian distribution always has a kurtosis of 3, it is convenient to define the excess kurtosis,$\gamma_{2}$, as
\begin{equation}
 \gamma_{2} = \frac{\mathbb{E}\left[ (x-\mu)^{4} \right] }{(\mathbb{E}\left[ (x-\mu)^{2} \right])^{2}} - 3 =\frac{m_{4}}{(m_{2})^{2}} - 3.
\end{equation} 
Using eq. 2 we find the fourth and second moments to be
\begin{align}
m_{4} &= \frac{1}{2}(\mu_{1}^{4} + 6\mu_{1}^{2}\sigma_{1}^{2} + 3\sigma_{1}^{4} + \mu_{2}^{4} + 6\mu_{2}^{2}\sigma_{2}^{2} + 3\sigma_{2}^{4}),  \\
m_{2} &= \frac{1}{2}(\mu_{1}^{2} + \sigma_{1}^{2} + \mu_{2}^{2} + \sigma_{2}^{2}).
\end{align}
Substituting eq. 4 and eq. 5 into eq. 3 we find the excess kurtosis of $P(x)$ to be given by
\begin{equation}
 \gamma_{2} = \frac{2(\mu_{1}^{4} + 6\mu_{1}^{2}\sigma_{1}^{2} + 3\sigma_{1}^{4} + \mu_{2}^{4} + 6\mu_{2}^{2}\sigma_{2}^{2} + 3\sigma_{2}^{4})}{(\mu_{1}^{2} + \sigma_{1}^{2} + \mu_{2}^{2} + \sigma_{2}^{2})^{2}} - 3.
\end{equation} 

In our system there is symmetry in the two modes about $x=0$. Therefore, we make the further assumption that $\mu_{2} = -\mu_{1}$, so that $\mu_{2}^{2} = \mu_{1}^{2} = \mu_{0}^{2}$ and let $\sigma_{1} = \sigma_{2} = \sigma_{0}$. $\gamma_{2}$ can then be written as

\begin{equation}
 \gamma_{2} = \frac{(2\mu_{0}^{4} + 12\mu_{0}^{2}\sigma_{0}^{2} + 6\sigma_{0}^{4})}{2(\mu_{0}^{2} + \sigma_{0}^{2})^{2}} - 3.
\end{equation} 

\section{Supplementary figures}

\begin{figure}[H]
	\centering
	\includegraphics[width=0.9\textwidth]{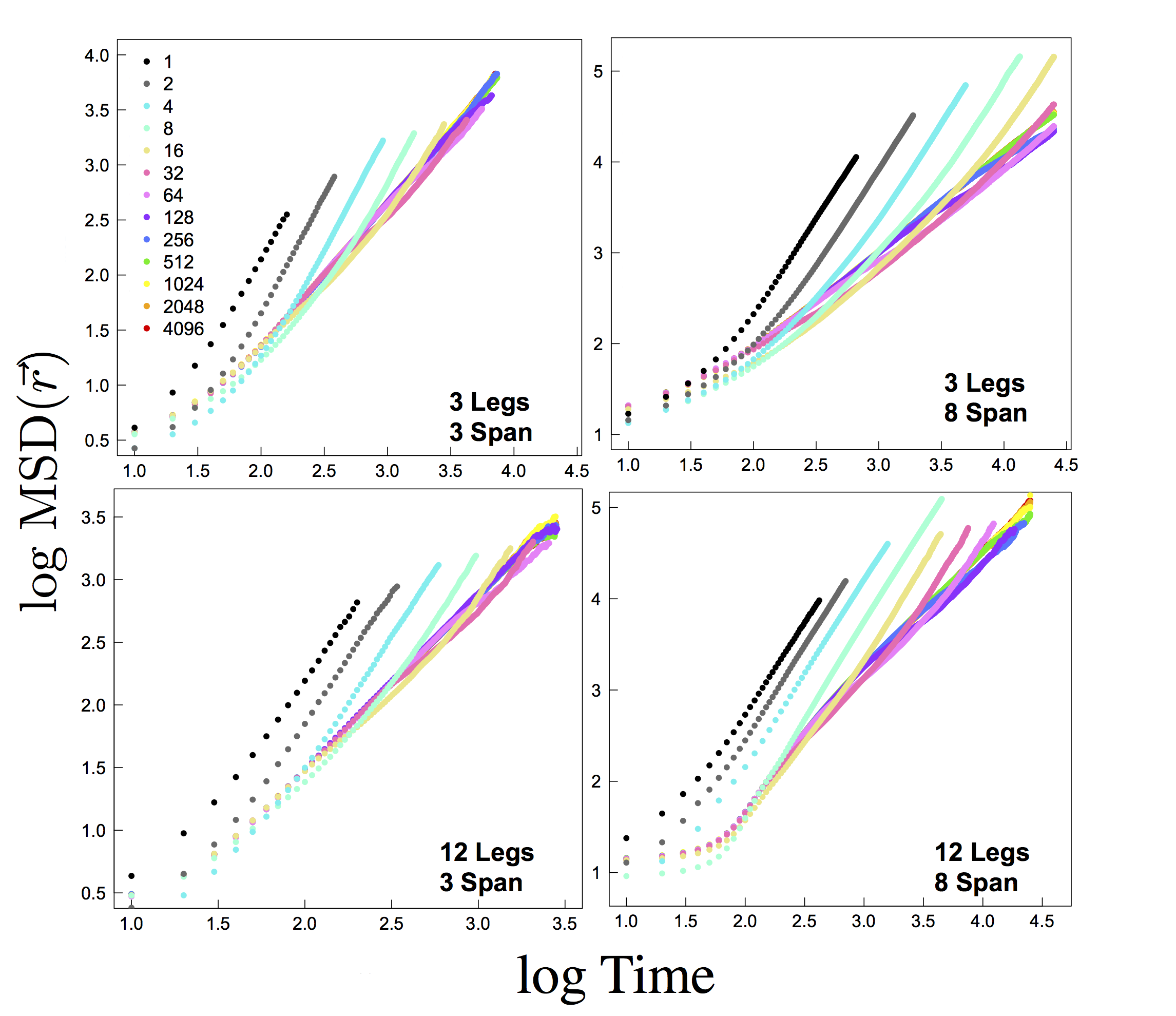}
	\caption{Each panel shows the evolution of MSD($\vec{r}$) for a given BBR design. MSD($\vec{r}$) for each of the BBR designs displays a similar trend with track width as discussed in the main text for (3,8) BBRs. The resulting $\alpha_{r}$ values from each MSD($\vec{r}$) curve are plotted in \textbf{Fig. 3b}.} 
\end{figure}

\begin{figure}[H]
	\centering
	\includegraphics[width=0.9\textwidth]{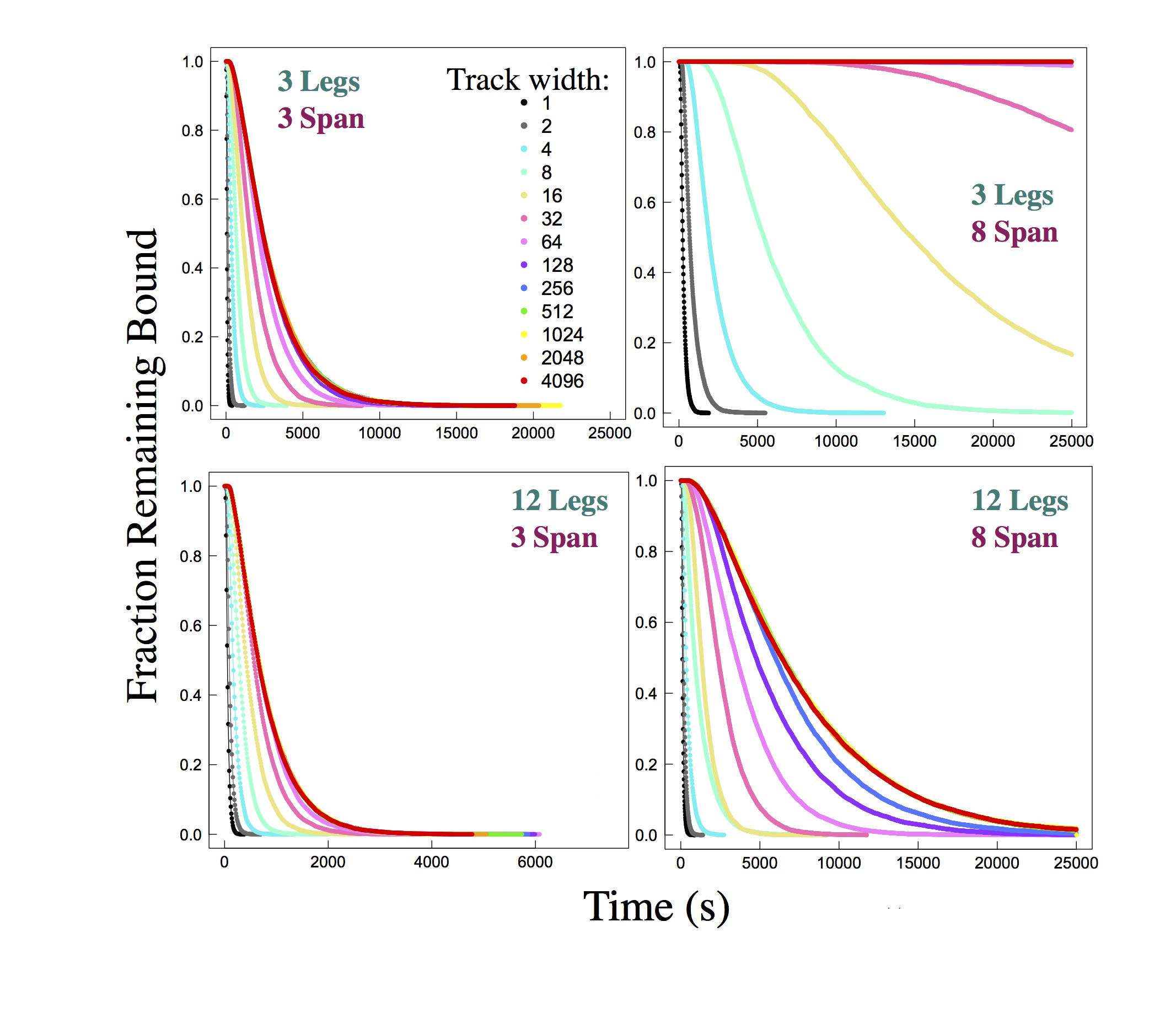}
	\caption{All detachment curves for each BBR design as a function of track width. (3,8) BBRs display negligible detachment for track widths larger than 32. All BBRs display a similar trend of longer track association as a function of increasing track width.} 
\end{figure}

\begin{figure}[H]
	\centering
	\includegraphics[width=0.9\textwidth]{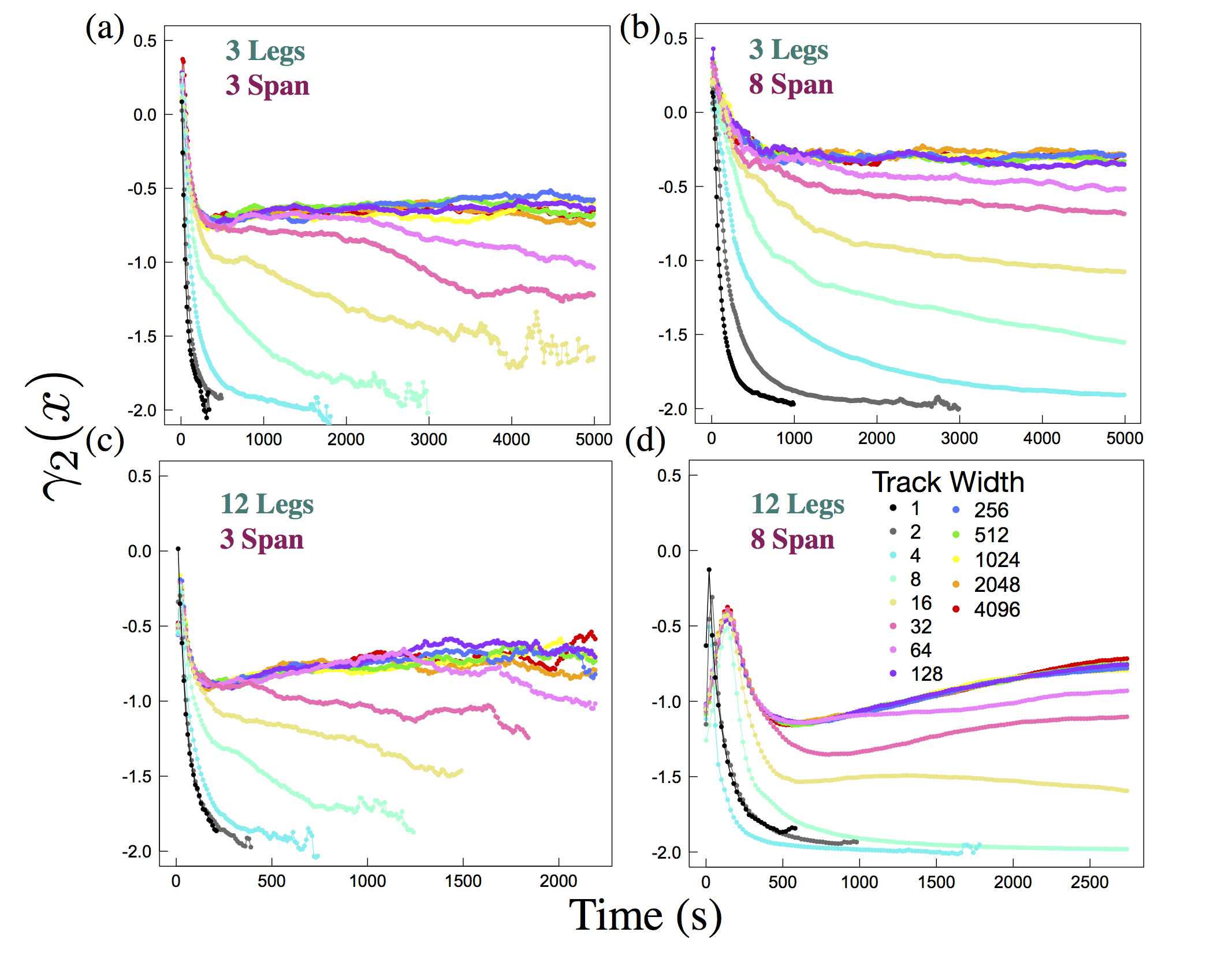}
	\caption{Excess kurtosis $\gamma_{2}(x)$, for all BBR designs across all track widths. a) (3,3) $\gamma_{2}(x)$ show similar trends to that of (3,8) BBRs, however, with their shorter track association times the curves are more noisy. b) $\gamma_{2}(x)$ for (3,8) BBRs, as shown and disucssed in the main text. c) (12,3) BBRs on narrow tracks display a decrease in $\gamma_{2}(x)$ to -2.0. However, on wider tracks and later times the BBRs experience an increase in $\gamma_{2}(x)$, indicating that over long times the dispersion increases at a greater rate than the modes separate. d) (12,8) BBRs display a similar trend in $\gamma_{2}(x)$ as (12,3) BBRs, where $\gamma_{2}(x)$ increases with time on wide tracks.} 
\end{figure}

\begin{figure}[H]
	\centering
	\includegraphics[width=0.9\textwidth]{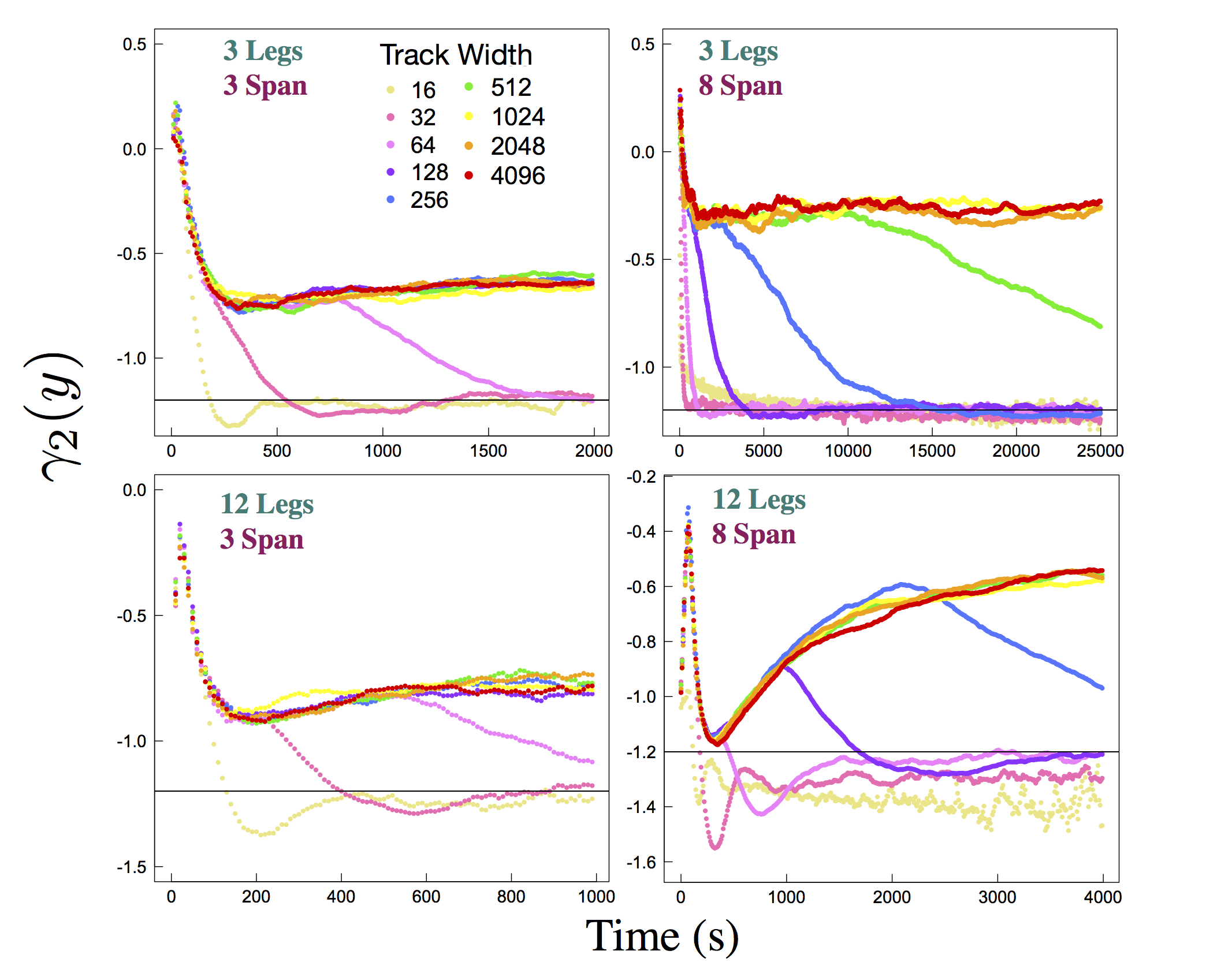}
	\caption{Excess kurtosis $\gamma_{2}(y)$, for all BBR designs across tracks of width 16-4096. For (3,3), (12,3), and (12,8) BBRs on intermediate tracks, $\gamma_{2}(y)$ decreases below -1.2 to a minimum before increasing to the -1.2 uniform distribution limit. (12,8) BBRs on wide tracks display an appreciable increase in $\gamma_{2}(y)$ over long times. For these BBRs, upon interacting with the confining walls $\gamma_{2}(y)$ turns over to a decreasing trend, as shown by the width 128 BBRs ((12,8), purple line). The $\gamma_{2}(y)$ trends for (3,8) BBRs are included and discussed in the main text.} 
\end{figure}

\begin{figure}[H]
	\centering
	\includegraphics[width=0.9\textwidth]{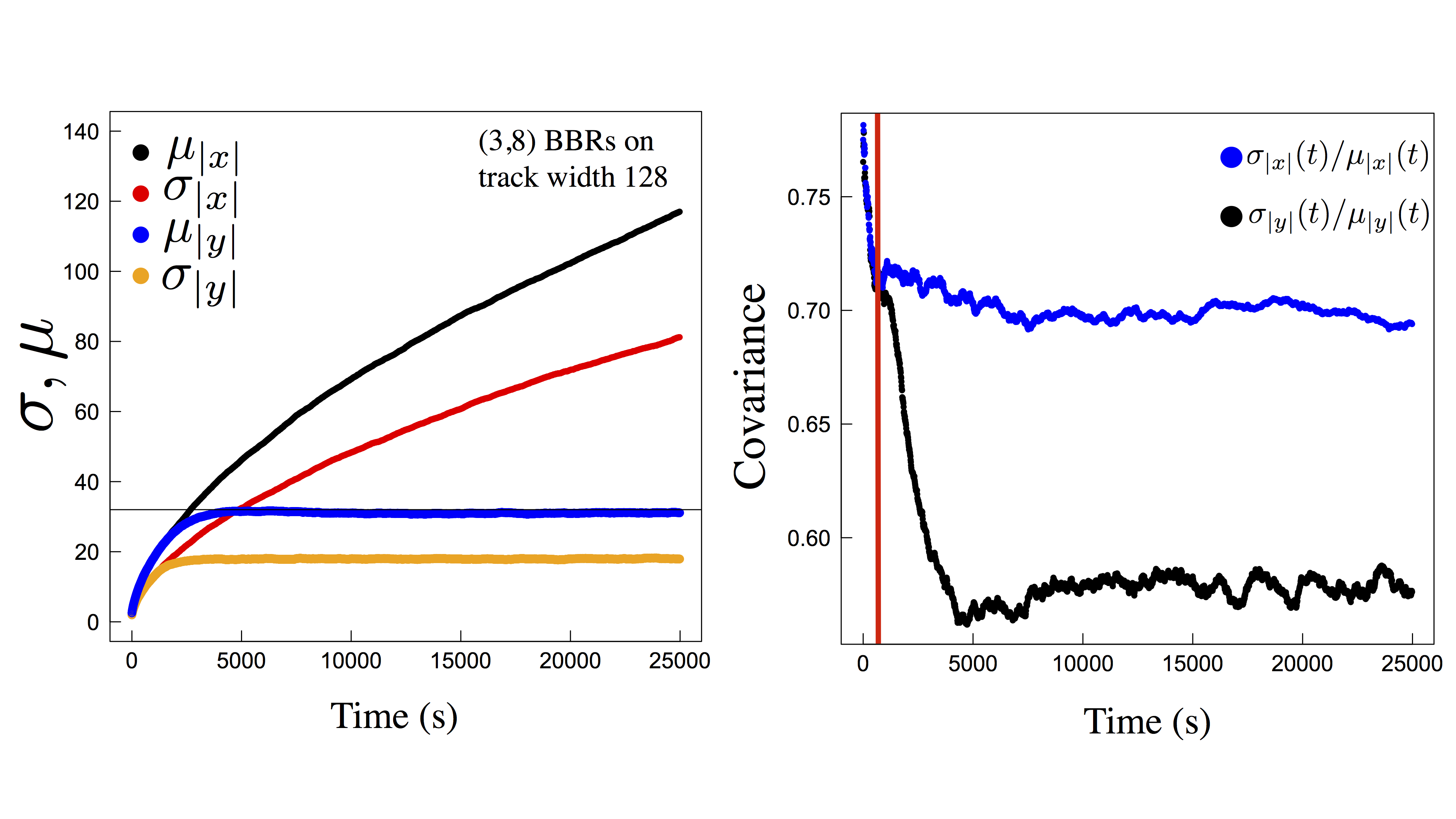}
	\caption{$\mu(t)$ and $\sigma(t)$ for BBRs with 3 legs, span 8 on a track width of 128. Left: $\mu_{\left| x \right| }(t)$ and $\mu_{\left| y \right|}(t)$ overlap until the ensemble of motors reaches the boundary at $\sim 1000s$, at which point $\mu_{\left| y \right| }(t)$ curves over to the expected value of 32 for a uniform distribution with boundaries at $|y| = 0,64$. Right: The ensemble average BBRs reach the boundary at $\sim 1000s$, indicated by the red line. $\frac{\sigma_{\left| x \right|}(t)}{\mu_{\left| x \right|}(t)}$ and $\frac{\sigma_{\left| y \right|}(t)}{\mu_{\left| y \right|}(t)}$, the coefficients of variation, indicate that the dispersion in \textit{y}-positions decreases substantially more than the mean $y$ position upon interaction with the boundary. $\frac{\sigma_{\left| x \right|}(t)}{\mu_{\left| x \right|}(t)}$ remains relatively constant following the onset of BBR interaction with the boundaries.} 
	\label{COV}
\end{figure}

\section{Supplementary movies}

\noindent{Movie S1: The ensemble position evolution for (3,3) BBRs on an effectively two-dimensional track. The ensemble of BBRs can be seen to develop a low occupancy around the centre (starting point) of the track.}\\
\\ 
\noindent{Movie S2: The ensemble position evolution for (3,8) BBRs on an effectively two-dimensional track. The ensemble of BBRs can be seen to maintain a high occupancy around the centre (starting point) of the track.}\\
\\
\noindent{Movie S3: The ensemble position evolution for (12,3) BBRs on an effectively two-dimensional track. The ensemble of BBRs can be seen to develop a low occupancy around the centre (starting point) of the track.}\\
\\ 
\noindent{Movie S4: The ensemble position evolution for (12,8) BBRs on an effectively two-dimensional track. The ensemble of BBRs can be seen to develop a low occupancy around the centre (starting point) of the track.}\\
\\ 
\noindent{Movie S5: The ensemble position evolution for (3,8) BBRs on a narrow track of width 8. Two modes can be seen to develop and move in opposite directions.} \\
\\
\noindent{Movie S6: The ensemble position evolution for (12,8) BBRs on a narrow track of width 8. Two modes can be seen to develop and move in opposite directions.}

%Design principles for synthetic molecular systems 
%Control over molecular 

%\bibliographystyle{ieeetr}
%\bibliography{Motor_Team.bib}